\documentclass[twocolumn,secnumarabic,amssymb, nobibnotes, aps, prd, superscriptaddress]{revtex4-2}

\usepackage{graphicx}
\usepackage{dcolumn}
\usepackage{bm}

\usepackage[utf8]{inputenc}
\usepackage[T1]{fontenc}
\usepackage{mathptmx}
\usepackage{etoolbox}
\usepackage{xcolor}

\usepackage{siunitx}
\usepackage{amsmath}

\begin{document}


\title{Magnetization fluctuations and magnetic aftereffect probed via the anomalous Hall effect} 



\author{Nadine Nabben}
\affiliation{Fachbereich Physik, Universität Konstanz, D-78457 Konstanz, Germany}

\author{Giacomo Sala}
\affiliation{ETH Zurich, Department of Materials, 8093 Zurich, Switzerland}

\author{Ulrich Nowak}
\affiliation{Fachbereich Physik, Universität Konstanz, D-78457 Konstanz, Germany}

\author{Matthias Krüger}
\affiliation{Institut für Theoretische Physik, Universität Göttingen, 37077 Göttingen, Germany}

\author{Sebastian T. B. Goennenwein}
\email[]{sebastian.goennenwein@uni-konstanz.de}
\affiliation{Fachbereich Physik, Universität Konstanz, D-78457 Konstanz, Germany}


\date{\today}

\begin{abstract}
Taking advantage of the anomalous Hall effect, we electrically probe low-frequency magnetization fluctuations at room temperature in a thin  ferromagnetic Pt/Co/AlO$_x$ layer stack with perpendicular magnetic anisotropy. We observe a strong enhancement of the Hall voltage fluctuations within the hysteretic region of the magnetization loop. 
Analyzing both the temporal evolution of the anomalous Hall voltage and its frequency-dependent noise power density, we identify two types of magnetic noise: abrupt changes in the magnetic domain configuration, evident as Barkhausen-like steps in the Hall voltage time trace, yield a noise power density spectrum scaling with frequency as $1/f^{\beta}$ with $\beta\approx 1.9$. In contrast, quasi-stationary magnetization configurations are connected with a magnetic noise power density with an exponent $\beta\approx 0.9$. The observation of Barkausen steps and relaxation effects shows that the magnetic system is in a non-stationary state in the hysteresis region, such that the fluctuation-dissipation theorem cannot be expected to hold. However, the time-dependent change in the Hall voltage for constant magnetic field strength resembles the integrated noise power.
\end{abstract}


\maketitle 

\section{Introduction}
Fluctuation phenomena limit the accuracy with which physical quantities can be measured. A detailed understanding of the mechanisms at play thus is important from a technical perspective. The fluctuations can furthermore provide interesting information about the physical systems themselves. For example, magnetic field sensors based on the anomalous Hall effect (AHE)\cite{nagaosa2010} exhibit significantly less noise at low frequencies as compared to sensors exploiting the ordinary Hall effect. Due to their metallic character, AHE sensors also are straightforward to fabricate and can have a broad frequency response up to multiple GHz\cite{zhang2019,zhang2020}. Vice versa, since the anomalous Hall voltage scales with the magnetization, AHE noise experiments are a powerful tool for studying magnetization fluctuations. Indeed, Diao et al.\cite{diao2010} observed Barkhausen jumps in AHE experiments in thin magnetic films, and invoked a mix of irreversible and reversible magnetic domain wall motion effects to explain the noise power with $1/f^{1.7}$ spectral dependence in the steep part of the magnetic hysteresis of their samples.

In magnetic thin films, different mechanisms have been proposed as sources for magnetization noise. One important mechanism is the so-called Barkhausen noise\cite{colaiori2008}.  
According to theory\cite{bertotti1981,bruno1990,Pommier-Barkhausen-2-processes-PRL65-1990,koch1993,Tadic-2000,kuntz2000}, both domain wall propagation and domain nucleation can result in Barkhausen-type effects. While abrupt, step-like, and typically irreversible changes in the magnetization are a characteristic fingerprint of Barkhausen processes, a smooth magnetization evolution is not sufficient to rule them out. In addition to Barkhausen effects, reversible excursions of domain walls has also been invoked as a mechanism for magnetic noise, in close analogy to a fluctuating two-level system \cite{dutta1981,diao2010}.  

Here, we take advantage of the anomalous Hall effect to analyze magnetization noise in magnetic thin film microstructures. In the magnetization hysteresis region, we observe AHE voltage noise with a power density that exhibits a $1/f^{\beta}$ shape with $\beta\approx 1.9$ if Barkhausen-like steps dominate the magnetization evolution with time. %
In contrast, the noise features an exponent $\beta\approx 0.9$ when the magnetization is quasi-stationary. 
Our data corroborate the experimental findings put forward by Diao et al.~\cite{diao2010}. 
However, a systematic analysis of the time evolution of the Hall voltage allows us to relate the different noise power exponents $\beta$ observed in experiment with the degree of stationarity of the magnetization state. While the fluctuation-dissipation theorem qualitatively reproduces the evolution of the noise power with magnetic field observed in experiment, we find that it quantitatively overestimates the noise in the hysteresis region. 
Furthermore, we observe that the time-dependent change of the anomalous Hall voltage for fixed magnetic field strength resembles the noise power. In other words, the occurrence of large magnetic noise power is directly connected with magnetic relaxation or magnetic after-effect\cite{Ferre-Dynamics-M-reversal-book-2002,magnetization-relaxation-review-Xi-Yang-JPCM20-2008}. The stationarity of the magnetic system must be carefully taken into account in the analysis of the noise response.

\section{Sample preparation and experimental methods}
\begin{figure}
    \includegraphics{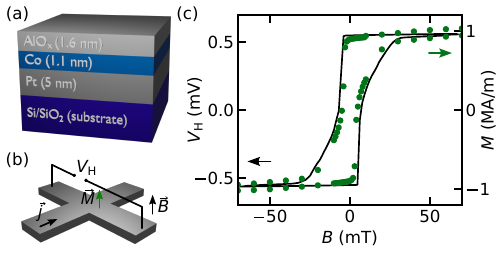}%
	  \caption{(a) schematic of the ferromagnetic layer stack. (b) schematic of the micropatterned Hall cross of the sample. The Hall voltage $V_\mathrm{H}$ is measured transversal to the current density $\vec{j}$. The external magnetic field $\vec{B}$ is applied perpendicular to the sample plane and therefore parallel to the easy axis of the thin film with perpendicular magnetic anisotropy.  (c) Hall voltage $V_\mathrm{H}$ (black line) measured while slowly ($\SI{1}{\minute/\milli\tesla}$) sweeping the external magnetic field. The green symbols show the magnetization $M$ acquired by SQUID magnetometry in another piece of the same sample. \label{fig:sample}}%
\end{figure}

For the investigation of magnetization fluctuations, a sample with large magnetic susceptibility and small active volume is most desirable\cite{diao2010}. We here use Pt($\SI{5}{\nano\meter}$)/Co($\SI{1.1}{\nano\meter}$)/AlO$_x$($\SI{1.6}{\nano\meter}$) thin films patterned into micron-scale Hall bars as schematically shown in Fig. \ref{fig:sample}(a) and (b). The layers are desposited on Si/SiO$_2$ substrates using sputter deposition. 
The microstructured Hall cross, as sketched in Fig. \ref{fig:sample}(b), has a sensitive area of $w_1 \times w_2 = 5\times 3 \,\si{\micro\meter}^2$, where $w_1$ describes the width of the current-carrying arm of the Hall cross and $w_2$ is the width of the transversal  voltage tap. Assuming that about $\SI{1}{\nano\meter}$ of the Pt layer also is magnetic owing to the proximity effect\cite{antel1999,wilhelm2000}, the active volume $Y$ relevant  in our Hall experiments thus is about $Y = \SI{32e-21}{\meter\tothe{3}}$. 

Our layer stack exhibits perpendicular magnetic anisotropy (PMA)\cite{carcia1988,rodmacq2009}, controllable via the AlO$_x$ layer thickness. The PMA is evident in the rather square hysteresis loop and small coercive fields observed with the external magnetic field $B=\mu_0 H$ applied perpendicular to the film plane, see Fig. \ref{fig:sample}(c). Note that we intentionally chose the  AlO$_x$ layer thickness such as to obtain a hysteresis loop with regions of different steepness, corresponding to different magnitudes of the magnetic susceptibility.  

Figure \ref{fig:sample}(c) also shows that the Hall voltage\cite{hurd1972} 
\begin{equation}
    V_\mathrm{H} = (R_0 ~\mu_0 H + R_\mathrm{s} ~\mu_0 M)\cdot I \label{eq:vh}, 
\end{equation}
with the vacuum permeability $\mu_0$, the ordinary Hall coefficient $R_0$ and the anomalous Hall coefficient $R_\mathrm{s}$ 
faithfully retraces the out-of-plane component of the magnetization $M$ of the sample  recorded in a separate experiment in a SQUID magnetometer on an unpatterned part of the same sample. 
The linear relation $V_\mathrm{H} \propto M$ (assumed to be instantaneous for the time scales considered here) is key for our  electrical investigation of magnetization noise. Assuming that fluctuations in the magnetic field and the drive current can be neglected, i.e., $\delta (\mu_0 H) = 0$ and $\delta I = 0$, Eq. \eqref{eq:vh} implies
\begin{equation}
    (\delta V_\mathrm{H})^2 \propto (\delta M)^2, 
\end{equation}
which justifies making statements about magnetization fluctuations while actually measuring fluctuations in the anomalous Hall voltage.\\
 
For the Hall voltage noise experiments, we apply a constant magnetic field of a given magnitude perpendicular to the sample plane and drive a dc current of $I= \SI{1}{\milli\ampere}$ along the Hall bar by connecting it to a battery with a suitable series resistor as a low-noise current source. 
For every given magnetic field magnitude, using a Zurich Instruments MFLI with MF-DIG option, we record the temporal evolution of the Hall voltage in fifty consecutive time periods of eighteen seconds each. Thus, $V_\mathrm{H}(t_i)$  corresponds to the voltage recorded in the time interval $t_i=[i\cdot \SI{18}{\second}, (i+1)\cdot \SI{18}{\second}]$, with $t_i$ the time after the magnetic field was changed.  
The noise power spectral density $S_{V,i}$ is obtained for each of these 50 $V_\mathrm{H}(t_i)$ time traces by making use of the Wiener-Khinchin theorem and calculating the Fourier transform of the autocorrelation function \cite{kogan2011}. Within each ($\SI{18}{\second}$ long) time trace, $2^{18}$ data points thus are recorded with a sample rate of $\SI{14.5}{\kilo\hertz}$. This results in a temporal resolution of $\Delta t = \SI{68}{\micro\second}$ and a resolution of the resulting noise spectrum of  $\Delta f = \SI{56}{\milli\hertz}$. 

\begin{figure}
    \includegraphics{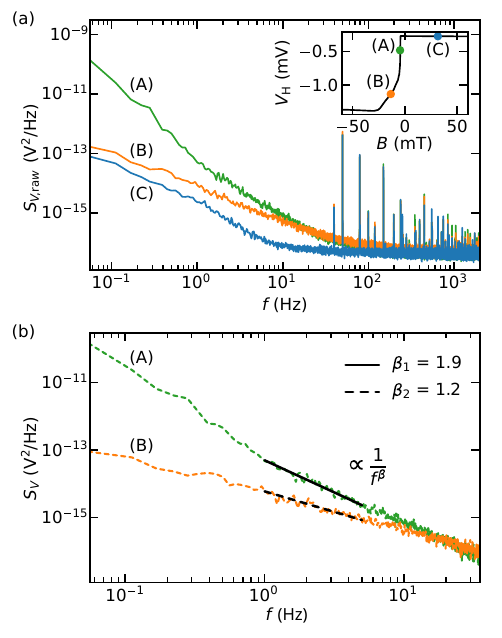}%
		\caption{(a) Noise power spectral density as calculated from temporal evolution of the Hall voltage recorded for different fixed external magnetic field strengths. The inset shows the Hall voltage $V_\mathrm{H}$ over the external magnetic field $B$. (b) Noise spectra as in (a), now with the background noise (spectrum (C), blue) subtracted. \label{fig:noise_raw}}%
\end{figure}

Figure \ref{fig:noise_raw}(a) shows three distinct noise spectra $S_{V,\mathrm{raw}}$ 
over the frequency $f$, each obtained by averaging a given set of 50 noise power density spectra $S_{V,i}$. 
The inset shows the Hall voltage $V_\mathrm{H}$ (black line) in dependence of the magnetic field and the colored points clarify at which magnetic field magnitude the corresponding noise spectrum is recorded.
All noise spectra show thermal Johnson-Nyquist noise\cite{johnson1928,nyquist1928,kogan2011} above $\SI{200}{\hertz}$ owing to the finite sample resistance. Even with good shielding and grounding, spikes at $f = \SI{37}{\hertz}$ and $f=\SI{50}{\hertz}$ and their harmonics are visible in the spectrum. Importantly, no such artefacts are present below $\SI{30}{\hertz}$. %
In some more detail, the noise spectrum recorded for saturated magnetization (spectrum (C), blue color in Fig. \ref{fig:noise_raw}(a)) shows only the background noise consisting of the inherent noise of the measurement electronics and the thermal Johnson-Nyquist noise of the sample. This background noise $S_{V,\mathrm{back}}$ is present in all noise spectra. 

To extract the magnetic noise, we subtract the noise spectrum for saturated magnetization as a background. The resulting $S_V = S_{V,\mathrm{raw}} - S_{V,\mathrm{back}}$ are shown in Fig. \ref{fig:noise_raw}(b). 
As evident from the figure and discussed in more detail in the next paragraphs, the magnitude and the spectral shape of the magnetic noise characteristically changes with the AHE slope or AHE susceptibility $\frac{\partial V_H}{\partial B}$. 
In the region of the hysteresis with a steeper slope, the noise is large and exhibits a spectral shape $1/f^{\beta}$ with $\beta\approx 1.9$ (curve (A) in Fig. \ref{fig:noise_raw}(b)). Koch \cite{koch1993} has argued that a continuous or step-like change of resistance yields a noise spectrum with an exponent $\beta=2$. Indeed, the temporal evolution of the Hall voltage in our samples shows discrete steps (Barkhausen steps) mostly in the region of the steep hysteresis. Koch's analysis thus corroborates the experimentally observed spectral shape with $\beta_1\approx 2$ for this magnetic field range.
When the external magnetic field is increased but the magnetization is not yet saturated, the measured noise is still well above the background noise. The noise power density observed in experiment then features $\beta_2\approx 1.2$ (curve (B) in Fig. \ref{fig:noise_raw}(b)). This is somewhat larger than $\beta=1$ expected for domain jittering\cite{dutta1981,diao2010}. However, as discussed in more detail in the context of Fig. \ref{fig:steps}, some Barkhausen-like features prevail also at larger field strengths, possibly resulting in $\beta_2 \approx 1.2 > 1$.

\section{Noise Analysis}

In an equilibrated system, the fluctuation-dissipation theorem (FDT) connects magnetic fluctuations with the imaginary or dissipative part of the magnetic susceptibility $\chi''_m(f)$ via\cite{hardner1993,kogan2011}
\begin{equation}
    S_m(f) = \frac{2k_\mathrm{B} T}{\pi \mu_0 f} \chi''_m(f) \label{eq:sm}.
\end{equation}
Here, $k_\mathrm{B}$ is the Boltzmann constant and $T$ the sample temperature. 
As detailed in \cite{kogan2011} and \cite{diao2010}, we reformulate Eq.~\eqref{eq:sm} from magnetic fluctuations to voltage fluctuations using 
$\chi''_V(f) =\frac{\mathrm{d}V_\mathrm{H}}{\mathrm{d} m} \chi''_m(f) $, $m=M\cdot Y$ and $S_V = ( \frac{\mathrm{d}V}{\mathrm{d}m} )^2 S_m$, such that
\begin{equation}
    S_V(f) = \frac{2k_\mathrm{B}T}{\pi f} \chi''_V(f) \frac{R_\mathrm{s} I}{Y}.
\end{equation}
The Kramers-Kronig relation connects the imaginary part $\chi''_V(f)$ with the real part of the susceptibility $\chi'_V$ at frequency $f=0$, such that the integrated noise power spectral density according to theory is:
\begin{equation}
     \int_0^{\infty} S_V(f) \mathrm{d}f = \frac{k_B T R_\mathrm{s} I \mu_0}{Y} \frac{\partial V_\mathrm{H}}{\partial B} = P_S^\mathrm{FDT} . \label{eq:theory}
\end{equation}

\begin{figure}[h!]
    \includegraphics{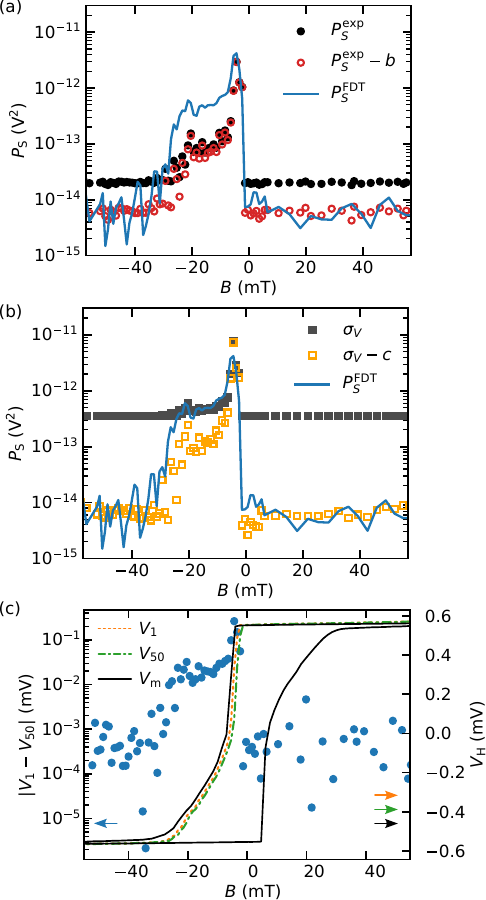}%
	  \caption{(a) The integrated noise power $P_S$ determined from experiment (full black circles) and the noise power $P_S^\mathrm{FDT}$ expected from the fluctuation-dissipation theorem (Eq.~\eqref{eq:theory}, blue line) exhibit a qualitatively similar magnetic field dependence. Subtraction of a constant offset $b$ from the experimental data however reveals quantitative differences (open red circles).
		(b) The variance calculated from the Hall voltage time traces (grey squares) also resembles the noise power $P_S^\mathrm{FDT}$ (blue line). After subtraction of constant offset $c$ (open orange squares), the discrepancy between experimental data and FDT prediction again becomes clearly apparent. 
		(c) The absolute change in Hall voltage during the measurement time for a given, constant external magnetic field point (blue points, left axis), also resembles  $P_S^\mathrm{FDT}$. The full lines show the Hall voltage $V_\mathrm{H}$ obtained from the first voltage time trace $V_1$ (orange dashed line, right axis) in the noise data taking, the last time trace $V_{50}$ (green dash-dotted line), and from the conventional magnetic field sweep with $\SI{1}{\minute/\milli\tesla}$ ($V_m$, black).  
    \label{fig:noise_int}}%
\end{figure}

Figure \ref{fig:noise_int}(a) shows the integrated noise power $P_S^\mathrm{exp} = \int_{\SI{56}{\milli\hertz}}^{\SI{30}{\hertz}} S_V(f) \mathrm{d}f$ (black circles) calculated from the measured raw noise power density, and the noise power $P_S^\mathrm{FDT}$ expected according to the fluctuation-dissipation theorem (Eq.~\eqref{eq:theory}, blue line). We account for amplifier noise and other magnetic-field independent noise contributions in the experimental data by subtracting a constant $b$ from $P_S^\mathrm{exp}$ resulting in the noise power shown by the red open circles. Hereby, $b$ is chosen such that the noise power in experiment matches the FDT prediction at high magnetic fields.	
Qualitatively, the evolution of the FDT prediction and the experimentally determined noise as a function of magnetic field are rather similar. However, their respective absolute values significantly differ in a large portion of the hysteresis region. 
Of course, the bounds of the integral in the experimental evaluation have a substantial influence on the magnitude of the noise power $P_S^\mathrm{exp}$, and the integration from $\SI{0}{\hertz}$ to $\infty$ required in Eq.~\eqref{eq:theory} cannot be experimentally realized. We here integrate the experimental data only for   $f\le\SI{30}{\hertz}$ in which the magnetic noise contributions dominate, while Johnson-Nyquist noise is prevalent at higher frequencies. The lower limit of the integration is determined by the minimal frequency $f=\SI{0.056}{\hertz}$ resolved in our experiment. 
However, while extending the integration bounds will certainly change the magnitude of noise power derived from experiment, it is difficult to rationalize that this will also remove the magnetic-field dependent difference between data and FDT prediction.   
Rather, we interpret the different noise power magnitudes in Fig.~\ref{fig:noise_int}(a) as indication that the FDT is not applicable to the hysteresis region, since the magnetic system is not stationary in this magnetic field range owing to Barkhausen-type processes and magnetic relaxation phenomena\cite{Ferre-Dynamics-M-reversal-book-2002,magnetization-relaxation-review-Xi-Yang-JPCM20-2008,diao2010}. 
In simple terms, in the hysteresis region the dissipation is enhanced relative to the fluctuations. 

In a complementary approach, we consider the variance $\sigma_V$, which in an ergodic stationary system is related to the integrated noise power density according to\cite{kogan2011}
\begin{equation}
    P_{S}^\mathrm{var}=  \int_0^{\infty} S_V(f) \mathrm{d}f = \overline{V_\mathrm{H}^2}-\overline{V_\mathrm{H}}^2=\sigma_V \label{eq:theorySTBG}
\end{equation}
The overbar here indicates a time average. Figure \ref{fig:noise_int}(b) shows that the variance $P_{S}^\mathrm{var}$ calculated from the experimental time traces (grey squares) is similar to $P_{S}^\mathrm{FDT}$ (blue line), at least within the magnetic hysteresis region. When the magnetization is saturated, however,  the value of $P_{S}^\mathrm{var}$ settles at a much higher base value. We attribute this to Johnson-Nyquist noise in the Hall voltage signal and the background noise of the measurement apparatus already mentioned above, which is inherently contained in the variance and cannot be suppressed by selecting appropriate parts of the frequency spectrum. Plotting $P_{S}^\mathrm{var} - c$ with the constant $c$ chosen to match $P_{S}^\mathrm{FDT}$ at high fields (orange open squares), we observe the same qualitative similarity but systematic quantitative differences between experimental data and FDT prediction as in panel (a).

The notion that large magnetic noise power arises in connection with non-stationary magnetic properties or finite so-called magnetic relaxation\cite{magnetization-relaxation-review-Xi-Yang-JPCM20-2008,Ferre-Dynamics-M-reversal-book-2002} is corroborated by Fig.~\ref{fig:noise_int}(c). In this figure, we depict $\left| V_1 - V_\mathrm{50} \right|$ as a function of the external magnetic field $B$ using full symbols. $V_1$ is the average of the first $\SI{18}{\second}$ long Hall voltage time trace, recorded after the magnetic field is set (see orange dotted line, right axis in Fig. \ref{fig:noise_int}(b)), while $V_\mathrm{50}$ is the average of the last time trace recorded about $\SI{15}{\minute}$ later (green dash-dotted line, right axis). For comparison, the $V_\mathrm{H}$-$B$ hysteresis loop, already shown in Fig. \ref{fig:sample}(c), taken with a sweep field rate of $\SI{1}{\minute/\milli\tesla}$ is depicted as black line. Figure \ref{fig:noise_int}(c) shows that the field sweep rate influences the shape of the Hall voltage hysteresis curve, and the finite value of $\left| V_1 - V_\mathrm{50} \right|$ demonstrates that in the hysteresis region, the Hall voltage is changing with time even for a constant applied magnetic field strength. The magnitude of this change is connected to the slope of the hysteresis. This appears reasonable, since the more susceptible the magnetic system is, the larger the time-dependent change or relaxation (at constant magnetic field) will be. 
Interestingly, the magnetic field-dependent magnitude of $\left| V_1 - V_\mathrm{50} \right|$ thus also resembles the integrated noise power over magnetic field depicted in panels (a) and (b). 

\begin{figure}
\includegraphics{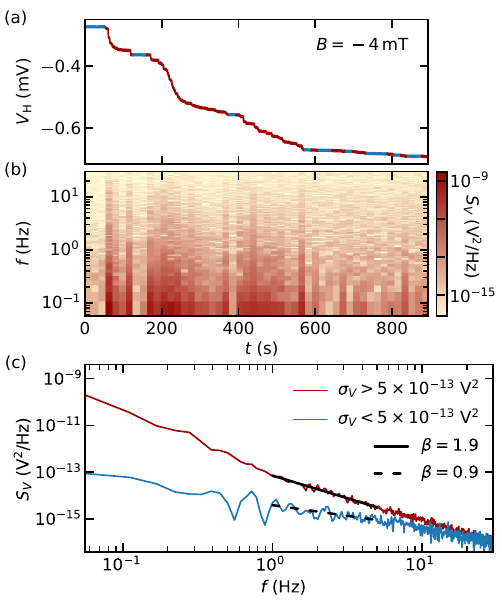}%
\caption{(a) Temporal evolution of $V_\mathrm{H}(t)$ at $B=\SI{-4}{\milli\tesla}$. (b) $S_V(f)$ of each $\SI{18}{\second}$ long slice of $V_\mathrm{H}(t_i)$ over frequency. (c) Averaged $S_V(f)$ if steps (no steps) are visible in $V_\mathrm{H}(t_i)$ in red (blue). The spectral shape if Barkhausen-like jumps (don't) occur is $f^{-1.9}$ ($f^{-0.9}$).  \label{fig:steps}}%
\end{figure}

Looking more closely at a typical temporal evolution of the Hall voltage for fixed magnetic field magnitude allows assessing the connection between step-like changes in the Hall voltage and the magnitude of the noise. Such a temporal evolution $V_\mathrm{H}(t)$ recorded at a constant external magnetic field $B = \SI{-4}{\milli\tesla}$ is shown in Fig.~\ref{fig:steps}(a).  
More specifically, the figure shows the sequence of 50 $V_\mathrm{H}(t_i)$ traces measured in the fashion discussed above in panel (a). For each repetition $i$, the noise power spectral density $S_V(f)$ is calculated and plotted in panel (b) over time - the time axis thus can also be seen as a reference to which $\SI{18}{\second}$-long time slice of $V_\mathrm{H}(t_i)$ the noise spectrum corresponds to. 
Comparing the shape of $V_\mathrm{H}(t)$ and the noise magnitude at low frequencies it is clearly evident that the noise is larger if more or larger steps occur in the temporal evolution of the Hall voltage. For example, in the first three sequences the magnetization of the sample (the Hall voltage) does not change much, resulting in low noise. Probably due to some thermal activation, the magnetization then starts to change, leading to an avalanche of step-like features in $V_\mathrm{H}(t)$ and much larger noise. In regions where no steps are visible (blue segments) but the magnetization is not saturated (e.g. at  $t = \SI{600}{\second}$) the noise power is smaller.

The false-color plot of Fig.~\ref{fig:steps}(b) provides direct proof that the noise power density qualitatively changes with time, for constant magnetic field. In order to correlate the spectral shape with the occurrence of steps in the time domain, we divide the noise spectra in two groups and average those groups separately. 
One group contains all spectra that show steps in the time domain (red line in Fig. \ref{fig:steps}(a)), the other all those that do not (blue line). 
We hereby assumed that a  variance $\sigma_V > \SI{5e-13}{\volt\squared}$ is indicative of a step (or step-like feature) in the respective time trace $i$.  
The averaged spectrum of each group is plotted over the frequency in \ref{fig:steps}(c). This figure shows that Barkhausen-like steps in the time domain yield noise with a spectral shape $1/f^{\beta}$ with $\beta \approx 1.9$. This is close to the prediction by Koch \cite{koch1993}, as already discussed in the context of Fig.~\ref{fig:noise_raw}. An exponent $\beta=2$ for Barkhausen noise is also predicted theoretically by Kuntz et al. by using mean field theory to calculate the power spectrum of an avalanche spin-flip\cite{kuntz2000}. 
In contrast, the absence of step-like features in the time domain results in  $1/f^{\beta}$ noise with an exponent $\beta\approx 0.9$. Since we subtracted the background noise from all spectra as discussed above, the fact that a clear noise signal prevails even in the absence of Barkhausen-like step features implies that another mechanism for magnetic noise is at play. 
Importantly, the two mechanisms apparently alternate as a function of time, for otherwise fixed experimental conditions. 
Simply calculating the noise power density from a given Hall voltage time trace, without carefully taking the time evolution or stationarity into account, is not sufficient for a robust analysis.

\section{Conclusion}
In summary, we probe magnetization fluctuations in a micropatterned magnetic thin film with perpendicular magnetic anisotropy by measuring the anomalous Hall effect response. We find that the magnitude of the noise power resembles the AHE voltage susceptibility $\partial V_\mathrm{H}/\partial B$. However, neither the integrated noise power $P_S$ nor the variance $\sigma_V$ of the Hall voltage quantitatively agree with the noise power calculated from the fluctuation-dissipation theorem. We attribute this to the fact that the system is in a non-stationary state in the hysteresis region, as evident from finite magnetic relaxation and Barkhausen-like steps in the time evolution of the Hall voltage. Interestingly, the magnitude of the time-dependent voltage change, $|V_1 - V_{50}|$, again resembles $\partial V_\mathrm{H}/\partial B$ or the magnetic noise power. 
Comparing the temporal evolution of the Hall voltage $V_\mathrm{H}(t)$ with the noise power spectral density in a series of time traces recorded at constant magnetic field magnitude, we find that abrupt changes in the magnetic state, visible as Barkhausen-like jumps in the time domain, result in large noise with a $1/f^{\beta}$ spectrum with $\beta\approx 1.9$. When the magnetization is quasi-stationary, we observe noise with a spectral shape of $1/f^{\beta}$ with $\beta\approx 0.9$. For a meaningful analysis, the non-stationary properties of the magnetization in the hysteresis region thus must be taken into account.

\begin{acknowledgments}
  We thank P. Gambardella, M. A. Weiss and X. Ai for fruitful discussions and input. 
	This work was financially supported by the Deutsche Forschungsgemeinschaft (DFG, German Research Foundation) 
	via the Collaborative Research Center SFB 1432 (Project no. 425217212).
\end{acknowledgments}

\bibliography{literature_AHEnoise}

\end{document}